\documentclass[a4paper,aps,twocolumn]{revtex4}
\begin{document}
\title{\bf{ Gravitational Analog of Faraday's Law via Torsion and a Metric with an Antisymmetric Part}}
\author{Philip~D.~Mannheim$^{1}$ and J.~J.~Poveromo$^{2}$}
\affiliation{$^1$Department of Physics, University of Connecticut, Storrs, CT 06269, USA.
email: philip.mannheim@uconn.edu\\
$^2$Department of Physics, Polytechnic Institute of New York University, Brooklyn, NY 11201, USA.
email: jp3800@nyu.edu}
\date{August 22, 2014}
\begin{abstract}
In this paper we show that in the presence of torsion and a metric with an antisymmetric part one can construct a gravitational analog of Faraday's law of electromagnetism.
\end{abstract}
\maketitle

\section{Introduction}

In a curved Riemannian background the covariant equations of motion for the electromagnetic  field $F_{\mu\nu}$ take the form
\begin{eqnarray}
\nabla_{\nu}F^{\nu\mu}=J^{\mu},
\label{1}
\end{eqnarray}
\begin{eqnarray}
&&(-g)^{-1/2}\epsilon^{\mu\nu\sigma\tau}\nabla_{\nu}F_{\sigma\tau}=0,
\label{2}
\end{eqnarray}
where $g$ is the determinant of the metric $g_{\mu\nu}$. Equation (\ref{2}) can also be written in the convenient form
\begin{eqnarray}
\nabla_{\nu}F_{\sigma\tau}+\nabla_{\tau}F_{\nu\sigma}+\nabla_{\sigma}F_{\tau\nu}=0,
\label{3}
\end{eqnarray}
and for brevity, we shall refer to Eq. (\ref{2}) as Faraday's Law even as it encompass Gauss' Law of Magnetism as well. Because of the antisymmetry of the  $\epsilon^{\mu\nu\sigma\tau}$ tensor density Levi-Civita symbol Eqs. (\ref{2}) and (\ref{3})  have no gravitational counterpart in a metric theory of gravity that is based on standard Riemannian geometry where both the metric and the connection have no antisymmetric piece. However, since the Cartan torsion tensor is antisymmetric on its last two indices, and since a 16-component metric tensor  would possess 6 antisymmetric components in addition to its standard 10 symmetric ones, with them one can introduce some antisymmetric structure into gravity theory. Here we show that precisely such antisymmetric structure will enable us to construct a gravitational analog of Faraday's Law. 

\section{The Levi-Civita, Torsion, and Spin Connections}

To establish our result we will need to consider the coupling of a fermion to  a geometry with torsion, and to this end we first recall how torsion is introduced into gravity theory. In order to construct covariant derivatives in any metric theory of gravity one must introduce a connection $\Gamma^{\lambda}_{\phantom{\alpha}\mu\nu}$, which has to transform under a coordinate transformation $x^{\mu}\rightarrow x^{\prime \mu}$ as 
\begin{eqnarray}
\Gamma^{\prime\lambda}_{\phantom{\alpha}\mu\nu}(x^{\prime})=
\frac{dx^{\prime\lambda}}{dx^{\alpha}}
\frac{dx^{\beta}}{dx^{\prime \mu}}
\frac{dx^{\gamma}}{dx^{\prime \nu}}
\Gamma^{\alpha}_{\phantom{\alpha}\beta\gamma}(x)+
\frac{d^2x^{\rho}}{dx^{\prime\mu}dx^{\prime \nu}}\frac{dx^{\prime\lambda}}{dx^{\rho}}.
\label{4}
\end{eqnarray}
For pure Riemannian geometry the connection is given by the Levi-Civita connection (here and throughout we follow the notation given in \cite{Fabbri2014})
\begin{eqnarray}
\Lambda^{\lambda}_{\phantom{\alpha}\mu\nu}=\frac{1}{2}g^{\lambda\alpha}(\partial_{\mu}g_{\nu\alpha} +\partial_{\nu}g_{\mu\alpha}-\partial_{\alpha}g_{\nu\mu}).
\label{5}
\end{eqnarray}
$\Lambda^{\lambda}_{\phantom{\alpha}\mu\nu}$ is symmetric in its $\mu$, $\nu$ indices, to thus have 40 independent components, and with it one can construct a covariant derivative operator $\nabla_{\mu}$, with the metric obeying  metricity conditions with indices sequenced here as
\begin{eqnarray}
\nonumber
&\nabla_{\mu}g^{\lambda\nu}=\partial_{\mu}g^{\lambda\nu}+\Lambda^{\lambda}_{\phantom{\alpha}\alpha\mu}g^{\alpha\nu}+\Lambda^{\nu}_{\phantom{\alpha}\alpha\mu}g^{\lambda\alpha}=0,\\
&\nabla_{\mu}g_{\lambda\nu}=\partial_{\mu}g_{\lambda\nu}-\Lambda^{\alpha}_{\phantom{\alpha}\lambda\mu}g_{\alpha\nu}-\Lambda^{\alpha}_{\phantom{\alpha}\nu\mu}g_{\lambda\alpha}=0.
\label{6}
\end{eqnarray}

To introduce torsion one takes the connection to no longer be symmetric on its two lower indices, and defines the Cartan torsion tensor $Q^{\lambda}_{\phantom{\alpha}\mu\nu}$ according to 
\begin{eqnarray}
Q^{\lambda}_{\phantom{\alpha}\mu\nu}=\Gamma^{\lambda}_{\phantom{\alpha}\mu\nu}-\Gamma^{\lambda}_{\phantom{\alpha}\nu\mu}.
\label{7}
\end{eqnarray}
With this antisymmetry $Q^{\lambda}_{\phantom{\alpha}\mu\nu}$ has 24 independent components. Unlike the Levi-Civita connection the torsion $Q^{\lambda}_{\phantom{\alpha}\mu\nu}$ transforms as a true rank three tensor under general coordinate transformations. In terms of the torsion tensor one defines a contorsion tensor according to 
\begin{eqnarray}
K^{\lambda}_{\phantom{\alpha}\mu\nu}=\frac{1}{2}g^{\lambda\alpha}(Q_{\mu\nu\alpha}+Q_{\nu\mu\alpha}-Q_{\alpha\nu\mu}),
\label{8}
\end{eqnarray}
and with $K^{\lambda}_{\phantom{\alpha}\mu\nu}$ one constructs the generalized connection
\begin{eqnarray}
\tilde{\Gamma}^{\lambda}_{\phantom{\alpha}\mu\nu}=\Lambda^{\lambda}_{\phantom{\alpha}\mu\nu}+K^{\lambda}_{\phantom{\alpha}\mu\nu}
\label{9},
\end{eqnarray}
to give a connection that now has 64 independent components, the maximum number possible in a 4-space. With this generalized connection also obeying Eq. (\ref{4}) (since $K^{\lambda}_{\phantom{\alpha}\mu\nu}$ transforms as a tensor), one can construct a covariant derivative operator $\tilde{\nabla}_{\mu}$, with the metric now obeying a generalized metricity condition
\begin{eqnarray}
\tilde{\nabla}_{\mu}g^{\lambda\nu}=\partial_{\mu}g^{\lambda\nu}+\tilde{\Gamma}^{\lambda}_{\phantom{\alpha}\alpha\mu}g^{\alpha\nu}+\tilde{\Gamma}^{\nu}_{\phantom{\alpha}\alpha\mu}g^{\lambda\alpha}=0
\label{10}
\end{eqnarray}
with respect to the connection $\tilde{\Gamma}^{\lambda}_{\phantom{\alpha}\mu\nu}$.

A torsion theory is then defined to be one in which one replaces $\Lambda^{\lambda}_{\phantom{\alpha}\mu\nu}$ by $\tilde{\Gamma}^{\lambda}_{\phantom{\alpha}\mu\nu}$, with the Riemann tensor 
\begin{eqnarray}
R^{\lambda}_{\phantom{\rho}\mu\nu\kappa}
=\partial_{\kappa}\Lambda^{\lambda}_{\phantom{\alpha}\mu\nu}-\partial_{\nu}\Lambda^{\lambda}_{\phantom{\alpha}\mu\kappa}
+\Lambda^{\eta}_{\phantom{\alpha}\mu\nu}\Lambda^{\lambda}_{\phantom{\alpha}\eta\kappa}
-\Lambda^{\eta}_{\phantom{\alpha}\mu\kappa}\Lambda^{\lambda}_{\phantom{\alpha}\eta\nu}~~
\label{11}
\end{eqnarray}
for instance being replaced by the Riemann-Cartan tensor
\begin{eqnarray}
\tilde{R}^{\lambda}_{\phantom{\rho}\mu\nu\kappa}
=\partial_{\kappa}\tilde{\Gamma}^{\lambda}_{\phantom{\alpha}\mu\nu}-\partial_{\nu}\tilde{\Gamma}^{\lambda}_{\phantom{\alpha}\mu\kappa}
+\tilde{\Gamma}^{\eta}_{\phantom{\alpha}\mu\nu}\tilde{\Gamma}^{\lambda}_{\phantom{\alpha}\eta\kappa}
-\tilde{\Gamma}^{\eta}_{\phantom{\alpha}\mu\kappa}\tilde{\Gamma}^{\lambda}_{\phantom{\alpha}\eta\nu}.~~
\label{12}
\end{eqnarray}

To incorporate half-integer spin first in a gravity theory without torsion one introduces a set of vierbeins $V^{a}_{\mu}$ in a torsionless Riemannian space where the coordinate $a$ refers to a fixed, special-relativistic reference coordinate system with metric $\eta_{ab}$, with the Riemannian metric then being writable as $g_{\mu\nu}=\eta_{ab}V^{a}_{\mu}V^{b}_{\nu}$. Because the vierbein carries a fixed basis index its covariant derivatives are not given via the Levi-Civita connection alone. Rather, one introduces an independent, and as yet torsion independent,  second connection known as the spin connection $\omega_{\mu}^{ab}$, with it being the derivative
\begin{eqnarray}
D_{\mu}V^{a\lambda}=\partial_{\mu}V^{a\lambda}+\Lambda^{\lambda}_{\phantom{\alpha}\nu\mu}V^{a \nu}+\omega_{\mu}^{ab}V^{\lambda}_{b}
\label{13}
\end{eqnarray}
that will transform as a tensor under both local translations and local Lorentz transformations provided the spin connection transforms as
\begin{eqnarray}
\omega_{\mu}^{\prime ab}=
\Lambda^a_{\phantom{a}c}(x)\Lambda^b_{\phantom{b}d}(x)\omega_{\mu}^{cd}
-\Lambda^{bc}(x)\partial_{\mu}\Lambda^a_{\phantom{a}c}(x)
\label{14}
\end{eqnarray}
under $V^a_{\mu}(x^{\lambda})\rightarrow \Lambda^a_{\phantom{a}c}(x)V^c_{\mu}(\Lambda^{\lambda}_{\phantom{\lambda}\tau}x^{\tau})$. If we now require metricity in the form $D_{\mu}V^{a\lambda}=0$, we find that $\omega_{\mu}^{ab}$ is then completely determined in terms of the vierbeins,  being given by the antisymmetric,  24-component 
\begin{eqnarray}
-\omega_{\mu}^{ab}=V^b_{\nu}\partial_{\mu}V^{a\nu}+V^b_{\lambda}\Lambda^{\lambda}_{\phantom{\lambda}\nu\mu}V^{a\nu}, 
\label{15}
\end{eqnarray}
i.e. by 
\begin{eqnarray}
-\omega_{\mu}^{ab}
&=&\frac{1}{2}(V^b_{\nu}\partial_{\mu}V^{a\nu}-V^a_{\nu}\partial_{\mu}V^{b\nu})
\nonumber\\
&+&\frac{1}{2}V^{b\alpha }V^{a\nu }(\partial_{\nu}g_{\alpha
\mu}-\partial_{\alpha}g_{\mu\nu})=\omega_{\mu}^{ba}.
\label{16}
\end{eqnarray}

To couple spinors to gravity in a Riemannian space without torsion, one starts with the free massless Dirac action in flat space, viz. the Poincare invariant $(1/2)\int d^4x \bar{\psi}\gamma^{a}i\partial_{a}\psi+H. c.$, where $\gamma_a\gamma_b+\gamma_b\gamma_a=2\eta_{ab}$. To make this action invariant under local translations one introduces a $(-g)^{1/2}$ factor in the measure and replaces $\gamma^{a}\partial_{a}$ by $\gamma^{a}V^{\mu}_a\partial_{\mu}$. While the resulting action is then invariant under spacetime independent Lorentz transformations of the form $\psi\rightarrow \exp(w^{ab}\Sigma_{ab})\psi$ where  $\Sigma_{ab}=(1/8)(\gamma_a\gamma_b-\gamma_b\gamma_a)$, when the function $w^{ab}$ is taken to be spacetime dependent, to continue to maintain invariance one has to augment the action with the spin connection of Eq. (\ref{16}), to then obtain the curved space Dirac action
\begin{eqnarray}
I_{\rm D}=\frac{1}{2}\int d^4x(-g)^{1/2}i\bar{\psi}\gamma^{a}V^{\mu}_a(\partial_{\mu}+\Sigma_{bc}\omega^{bc}_{\mu})\psi +H. c.~~
\label{17}
\end{eqnarray}
An integration by parts in $I_{\rm D}$ and use of 
\begin{eqnarray}
&&\partial_{\mu}V^{a\mu}+(-g)^{-1/2}\partial_{\mu}(-g)^{1/2}V^{a\mu}+\omega_{\mu}^{ab}V^{\mu}_{b}=0,
\nonumber\\
&&\gamma^a[\gamma^b,\gamma^c]-[\gamma^b,\gamma^c]\gamma^{a}=4\eta^{ab}\gamma^c-4\eta^{ac}\gamma^b
\label{18}
\end{eqnarray} 

yields 
\begin{eqnarray}
I_{\rm D}&=&\int d^4x(-g)^{1/2}i\bar{\psi}\gamma^{a}V^{\mu}_a(\partial_{\mu}+\Sigma_{bc}\omega^{bc}_{\mu})\psi
\nonumber\\
&-&\frac{1}{2}\int d^4x(-g)^{1/2}i\bar{\psi}\gamma_{a}V^{\mu}_b\omega^{ab}_{\mu}\psi
\nonumber\\
&+&\frac{1}{2}\int d^4x(-g)^{1/2}i\bar{\psi}V^{a \mu}(\Sigma_{bc}\gamma_{a}-\gamma_{a}\Sigma_{bc})\omega^{bc}_{\mu}\psi
\nonumber\\
&=&\int d^4x(-g)^{1/2}i\bar{\psi}\gamma^{a}V^{\mu}_a(\partial_{\mu}+\Sigma_{bc}\omega^{bc}_{\mu})\psi.
\label{19}
\end{eqnarray}

To now generalize the spin connection and the Dirac equation to a space with torsion one replaces $\Lambda^{\lambda}_{\phantom{\alpha}\mu\nu}$ by $\tilde{\Gamma}^{\lambda}_{\phantom{\alpha}\mu\nu}$, with the spin connection and metricity conditions generalizing to 
\begin{eqnarray}
\nonumber
\tilde{D}_{\mu}V^{a\lambda}&=&\partial_{\mu}V^{a\lambda}+(\Lambda^{\lambda}_{\phantom{\alpha}\nu\mu}+K^{\lambda}_{\phantom{\alpha}\nu\mu})V^{a \nu}+\tilde{\omega}_{\mu}^{ab}V^{\lambda}_{b}=0
,\\
-\tilde{\omega}_{\mu}^{ab}&=&-\omega_{\mu}^{ab}+V^{b}_{\lambda}K^{\lambda}_{\phantom{\alpha}\nu\mu}V^{a \nu}=\tilde{\omega}_{\mu}^{ba}.
\label{20}
\end{eqnarray}
With this $\tilde{\omega}_{\mu}^{ab}$, we obtain a torsion-dependent Dirac action of the form
\begin{eqnarray}
\tilde{I}_{\rm D}=\frac{1}{2}\int d^4x(-g)^{1/2}i\bar{\psi}\gamma^{a}V^{\mu}_a(\partial_{\mu}+\Sigma_{bc}\tilde{\omega}^{bc}_{\mu})\psi+H. c.~~
\label{21}
\end{eqnarray}
An integration by parts in $\tilde{I}_{\rm D}$ yields
\begin{eqnarray}
\tilde{I}_{\rm D}&=&I_{\rm D}+\frac{1}{16}\int d^4x(-g)^{1/2}i\bar{\psi}V^{a\mu}V^{b}_{\lambda}V^{c \nu}
\nonumber\\
&\times&(K^{\lambda}_{\phantom{\alpha}\nu\mu}\gamma_a[\gamma_b,\gamma_c]+
(K^{\lambda}_{\phantom{\alpha}\nu\mu})^{\dagger}[\gamma_b,\gamma_c]\gamma_a)\psi,
\label{22}
\end{eqnarray}
where $I_{\rm D}$ is the torsion-independent Dirac action given in Eq. (\ref{19}). On taking $K^{\lambda}_{\phantom{\alpha}\nu\mu}$ to be Hermitian (i.e. on taking $Q^{\lambda}_{\phantom{\alpha}\nu\mu}$ to be real), use of  
\begin{eqnarray}
\gamma^a[\gamma^b,\gamma^c]+[\gamma^b,\gamma^c]\gamma^a&=&4i\epsilon^{abcd}\gamma_{d}\gamma^{5}, 
\nonumber\\
 \gamma^5&=&i\gamma^0\gamma^1\gamma^2\gamma^3,
\nonumber\\
\epsilon^{abcd}V^{\mu}_aV^{\nu}_bV^{\sigma}_cV^{\tau}_d&=&(-g)^{-1/2}\epsilon^{\mu\nu\sigma\tau}
\label{23}
\end{eqnarray}
yields \cite{Shapiro2002}
\begin{eqnarray}
\tilde{I}_{\rm D}=\int d^4x(-g)^{1/2}i\bar{\psi}\gamma^{a}V^{\mu}_a(\partial_{\mu}+\Sigma_{bc}\omega^{bc}_{\mu}-i\gamma^5S_{\mu})\psi,~~
\label{24}
\end{eqnarray}
where 
\begin{eqnarray}
S^{\mu}&=&\frac{1}{8}(-g)^{-1/2}\epsilon^{\mu\alpha\beta\gamma}Q_{\alpha\beta\gamma}.
\label{25}
\end{eqnarray}
Recalling that $\epsilon^{0123}=1$, $\epsilon_{0123}=-1$, use of
\begin{eqnarray}
&&-(-g)^{-1}\epsilon_{\mu\nu\sigma\tau}\epsilon^{\mu\alpha\beta\gamma}=
\delta^{\alpha}_{\nu}\delta^{\beta}_{\sigma}\delta^{\gamma}_{\tau}+
\delta^{\alpha}_{\tau}\delta^{\beta}_{\nu}\delta^{\gamma}_{\sigma}+
\delta^{\alpha}_{\sigma}\delta^{\beta}_{\tau}\delta^{\gamma}_{\nu}
\nonumber\\
&&\qquad
-\delta^{\alpha}_{\nu}\delta^{\beta}_{\tau}\delta^{\gamma}_{\sigma}-
\delta^{\alpha}_{\sigma}\delta^{\beta}_{\nu}\delta^{\gamma}_{\tau}-
\delta^{\alpha}_{\tau}\delta^{\beta}_{\sigma}\delta^{\gamma}_{\nu}
\label{26}
\end{eqnarray}
enables us to rewrite Eq. (\ref{25}) in the form
\begin{eqnarray}
-(-g)^{-1/2}\epsilon_{\alpha\beta\gamma\mu}S^{\mu}=\frac{1}{4}[Q_{\alpha\beta\gamma}+Q_{\gamma\alpha\beta}+Q_{\beta\gamma\alpha}].
\label{27}
\end{eqnarray}

In the action $\tilde{I}_{\rm D}$  we note that even though the torsion is only antisymmetric on two of its indices, the only components of the torsion that appear in its torsion-dependent $S^{\mu}$ term are the four that constitute that part of the torsion that is antisymmetric on all three of its indices. 

With $\tilde{I}_{\rm D}$ admitting of an immediate generalization that incorporates electromagnetism, viz. 
\begin{eqnarray}
\tilde{I}_{\rm D}&=&\int d^4x(-g)^{1/2}i\bar{\psi}\gamma^{a}V^{\mu}_a(\partial_{\mu}+\Sigma_{bc}\omega^{bc}_{\mu}
\nonumber\\
&-&iA_{\mu}-i\gamma^5S_{\mu})\psi,
\label{28}
\end{eqnarray}
we see that in Eq. (\ref{28}) $S_{\mu}$ is an axial field  that minimally couples to the axial fermion current in exactly the same minimally coupled way as the  standard electromagnetic vector potential $A_{\mu}$ couples to the vector fermion current. Because of this minimal coupling $\tilde{I}_{\rm D}$ is both locally gauge invariant under $\psi\rightarrow e^{i\alpha(x)}\psi$, $A_{\mu}\rightarrow A_{\mu}+\partial_{\mu}\alpha(x)$, and locally chiral invariant \cite{Shapiro2002} under $\psi\rightarrow e^{i\gamma^5\beta(x)}\psi$, $S_{\mu}\rightarrow S_{\mu}+\partial_{\mu}\beta(x)$. Additionally, as noted in \cite{Fabbri2014}, $\tilde{I}_{\rm D}$ is locally conformal invariant under $V^a_{\mu}(x)\rightarrow \Omega(x)V^a_{\mu}(x)$, $\psi(x)\rightarrow \Omega^{-3/2}(x)\psi(x)$ since, just like the vector potential $A_{\mu}$, the equally minimally coupled $S_{\mu}$ also has zero conformal weight \cite{footnote1}. The $\tilde{I}_{\rm D}$ action thus has a remarkably rich local invariance structure, as it is invariant under local translations, local Lorentz transformations, local gauge transformations, local axial transformations, and local conformal transformations. 

\section{Gravitational Analog of Faraday's Law}

We note immediately the similarity between the structure of the gravitational Eqs. (\ref{25}) and (\ref{27}) on the one hand and the electromagnetic Eqs. (\ref{2}) and (\ref{3}) on the other, with both involving a rank three tensor (viz. $\nabla_{\mu}F_{\nu\tau}$ and $Q_{\mu\nu\tau}$) that is antisymmetric on its last two indices. To make the analog complete, we need to posit  that the torsion can be written as the covariant derivative of some, for the moment, general rank two tensor $A_{\mu\nu}$ that is itself antisymmetric \cite{footnote2}. Thus we posit that one can write the torsion tensor as 
\begin{eqnarray}
Q_{\mu\nu\tau}=\nabla_{\mu}A_{\nu\tau},
\label{29}
\end{eqnarray}
where $\nabla_{\mu}$ denotes the Levi-Civita-based covariant derivative  and not the generalized torsion-based covariant derivative $\tilde{\nabla}_{\mu}$. With Eq. (\ref{29}) we then obtain a gravitational analog of the electromagnetic Faraday Law, viz.  
\begin{eqnarray}
\nabla_{\nu}A_{\sigma\tau}+\nabla_{\tau}A_{\nu\sigma}+\nabla_{\sigma}A_{\tau\nu}=-4(-g)^{1/2}\epsilon_{\nu\sigma\tau\mu}S^{\mu},
\label{30}
\end{eqnarray}
with $A_{\mu\nu}$ serving as a gravitational analog of the electromagnetic $F_{\mu\nu}$, and with $S^{\mu}$ serving as a gravitational analog of a magnetic monopole current should Faraday's Law possess such a source. Finally,  if we set $S^{\mu}$ equal to zero, we then obtain a Faraday-type law for gravity of the form 
\begin{eqnarray}
\nabla_{\nu}A_{\sigma\tau}+\nabla_{\tau}A_{\nu\sigma}+\nabla_{\sigma}A_{\tau\nu}=0,
\label{31}
\end{eqnarray}
to thus establish that  via the coupling of torsion to fermions it is in principle possible to construct a gravitational analog of Faraday's Law of electromagnetism.

\section{Physical Significance of $A_{\mu\nu}$}

While we have introduced $A_{\mu\nu}$ solely as a general antisymmetric rank two curved space tensor that is  related to the torsion tensor via $Q_{\mu\nu\tau}=\nabla_{\mu}A_{\nu\tau}$, it is of interest to ask whether, given its antisymmetry, $A_{\mu\nu}$ could be identified with some specific antisymmetric rank two tensor that has previously been encountered in the literature. To this end we note that there are three physical possibilities for $A_{\mu\nu}$ that immediately come to mind, namely that $A_{\mu\nu}$ be identified with the electromagnetic Maxwell tensor $F_{\mu\nu}$, that it be identified with the Kalb-Ramond field $B_{\mu\nu}$ of string theory,  or that it be identified with the antisymmetric part of what would then be a 16-component metric tensor. 

In the past there has been much study in the literature of the possibility that the electromagnetic $F_{\mu\nu}$ could have a geometric origin, as it could potentially lead to a unification of gravitation and electromagnetism \cite{Einstein1946}. However, an identification of the $A_{\mu\nu}$ we have introduced with $F_{\mu\nu}$ cannot be immediate since not only do these two tensors have different engineering dimension ($A_{\mu\nu}$ is dimensionless while  $F_{\mu\nu}$ has dimension of inverse length squared), they also have different conformal weights ($A_{\mu\nu}$ has conformal weight two while  $F_{\mu\nu}$ has conformal weight zero). Thus to make an identification of $A_{\mu\nu}$ with $F_{\mu\nu}$ one would have to introduce a parameter such as Newton's constant with the dimension of length squared. However that would not be compatible with the conformal invariance of the $\tilde{I}_{\rm D}$ action from whence we obtained $S^{\mu}$ in the first place, and so  we do not consider this possibility further.

A second possibility for $A_{\mu\nu}$ is that it would be identified with the Kalb-Ramond field, $B_{\mu\nu}$, a massless excitation in string theory \cite{Kalb1974}. This $B_{\mu\nu}$ field serves as a generalization of the electromagnetic vector potential $A_{\mu}$ to the string theory case, with $B_{\mu\nu}$ coupling to strings in a manner analogous to the way $A_{\mu}$ couples to point particles. In certain formulations of string theory $B_{\mu\nu}$ accompanies the graviton field that is associated with the symmetric part of the metric. Moreover, associated with $B_{\mu\nu}$ is a rank three field strength generalization of the rank two Maxwell tensor $F_{\mu\nu}$,  viz. $H_{\mu\lambda\nu}=\partial_{\mu}A_{\lambda\nu}+\partial_{\nu}A_{\mu\lambda}+\partial_{\lambda}A_{\nu\mu}$. Just like the quantity $Q_{\alpha\beta\gamma}+Q_{\gamma\alpha\beta}+Q_{\beta\gamma\alpha}$ that appears on the right-hand side of Eq. (\ref{27}), $H_{\mu\lambda\nu}$ is also antisymmetric on all three of its indices. It would thus be of interest to see if one could relate $H_{\mu\lambda\nu}$ to $Q_{\mu\lambda\nu}$, though that would be beyond the scope of this paper.
 
As to our third possibility, an identification of $A_{\mu\nu}$ with the antisymmetric part of the spacetime metric is actually quite feasible since, like $A_{\mu\nu}$, the metric also has zero engineering dimension and conformal weight two. Moreover, with such an identification we can then construct a geometrical analog of Faraday's Law that fully respects the very same local conformal invariance that the electromagnetic Maxwell equations themselves possess \cite{footnote3}. And as we now show, in a non-symmetric gravity theory it is actually possible to make an identification of $A_{\mu\nu}$ with the antisymmetric part of the metric just as we would want.

Now taking the spacetime metric to have a non-vanishing antisymmetric part would constitute a modification of standard, symmetric metric  tensor gravity, and such a modification has been studied in and of itself quite extensively in the literature \cite{Moffat1979,Moffat1980,Moffat1995,Janssen2006,Poplawski2007}. Moreover, if we do wish to identify $A_{\mu\nu}$ with the antisymmetric part of the spacetime metric we would need to enlarge the theory we have developed above and embed it in a theory that is built on a 16-component metric tensor from the very beginning. Various approaches have been developed in the literature to treat such a 16-component metric theory, since once the metric is not symmetric one can define differing covariant derivatives depending on the sequencing of indices. Moreover, one can either specify the connection to be the one that makes some specific covariant derivative of the metric zero, or one can leave the connection unspecified a priori and allow the equations of motion themselves to determine the connection by stationary functional variation with respect to it (Palatini approach). In this latter approach the resulting variation is not guaranteed to lead to a vanishing covariant derivative of the metric, and in general even in the symmetric metric case does not do so for actions other than the Einstein-Hilbert action itself. 

Of particular interest for our purposes here is the specific non-symmetric gravity theory developed by Moffat. Moffat's approach  is based on the Einstein-Hilbert action, and using  the Palatini formalism he found \cite{Moffat1980} that in a weak gravity expansion around a flat Minkowski metric $\eta_{\mu\nu}$, the first-order term in the connection takes the form
\begin{eqnarray}
\Gamma^{\lambda}_{\phantom{\alpha}\mu\nu}&=&\frac{1}{2}\eta^{\lambda\alpha}(\partial_{\mu}g^{\rm S}_{\alpha\nu} +\partial_{\nu}g^{\rm S}_{\mu\alpha}-\partial_{\alpha}g^{\rm S}_{\nu\mu})
\nonumber\\
&+&\frac{1}{2}\eta^{\lambda\alpha}(\partial_{\mu}g^{\rm A}_{\alpha\nu} +\partial_{\nu}g^{\rm A}_{\mu\alpha}-\partial_{\alpha}g^{\rm A}_{\nu\mu}),
\label{32}
\end{eqnarray}
where $g^{\rm S}_{\mu\nu}$ is the symmetric part of the 16-component metric tensor and $g^{\rm A}_{\mu\nu}$ is its antisymmetric part. On comparing Eq. (\ref{32}) with Eq. (\ref{9}) we see that the antisymmetric part of the connection is written as none other the derivative of the antisymmetric part of the metric just as we want \cite{footnote4}. Thus as we see, in non-symmetric gravity theory one is able to derive a relation between the torsion and the derivative of the antisymmetric part of the spacetime metric, and in so doing one is then able to derive a gravitational analog of Faraday's Law of electromagnetism.


\begin{thebibliography}{9}
%
\bibitem{Fabbri2014} 
L.~Fabbri and P.~D.~Mannheim,~Phys.~Rev.~D~\textbf{90}, 024042 (2014).
%
%
\bibitem{Shapiro2002}
I.~L.~Shapiro,~Phys.~Rept.~\textbf{357}, 113 (2002).
%
%
\bibitem{footnote1} Under a local conformal transformation the torsion transforms as \cite{Shapiro2002}  $Q^{\lambda}_{\phantom{\lambda}\mu\nu}\rightarrow Q^{\lambda}_{\phantom{\lambda}\mu\nu}+q\Omega^{-1}(x)(\delta^{\lambda}_{\mu}\partial_{\nu}-\delta^{\lambda}_{\nu}\partial_{\mu})\Omega(x)$ where $q$ is its conformal weight. No matter what the value of $q$, under a conformal transformation $S_{\mu}$ transforms into itself with all derivatives of $\Omega(x)$ dropping out.
%
\bibitem{footnote2} We are not of course asserting that $Q_{\mu\nu\tau}$ necessarily is given by the covariant derivative of an antisymmetric rank two tensor, but only noting that if it is, one is then able to construct a geometrical  analog of Faraday's Law.
%
\bibitem{Einstein1946} A.~Einstein and E.~G.~Straus,~Ann.~Math.~\textbf{47}, 731 (1946).
%
\bibitem{Kalb1974} M.~Kalb and P.~Ramond~Phys.~Rev.~D \textbf{9}, 2273 (1974).
%
\bibitem{footnote3} For a possible role for torsion in Faraday's Law of electromagnetism itself see P.~D.~Mannheim, \textit{Torsion, Magnetic Monopoles and Faraday's Law  via a Variational Principle},~arXiv:1406.2265 [hep-th], June, 2014; P.~D.~Mannheim, \textit{$PT$ Symmetry, Conformal Symmetry, and the Metrication of Electromagnetism},~arXiv:1407.1820 [hep-th], July, 2014.
%
\bibitem{Moffat1979} J.~W.~Moffat,~Phys.~Rev.~D~\textbf{19}, 3554 (1979).
%
\bibitem{Moffat1980} J.~W.~Moffat,~J.~Math.~Phys.~\textbf{21}, 1798 (1980).
%
\bibitem{Moffat1995} J.~W.~Moffat,~Phys.~Lett.~B~\textbf{355}, 447 (1995).
%
\bibitem{Janssen2006} T.~Janssen and T.~Prokopec,~Class.~Quant.~Grav.~\textbf{23}, 4967 (2006).
%
\bibitem{Poplawski2007} N.~J.~Poplawski,~Mod.~Phys.~Lett.~A \textbf{22}, 2701 (2007). 
%
\bibitem{footnote4} In regard to Eq. ({32}) we should note that since the Maxwell equations are linear while gravitational equations are nonlinear, we might expect a relation such as $Q_{\mu\nu\tau}=\nabla_{\mu}A_{\nu\tau}$ to only hold perturbatively anyway.
%
\end{thebibliography}
\end{document}